\documentclass[prb,showpacs,preprintnumbers,amsmath,amssymb,twocolumn,superscriptaddress,longbibliography]{revtex4-1}
\usepackage{graphicx}
\usepackage{gensymb}
\def\be{\begin{equation}}
\def\ee{\end{equation}}
\def\bea{\begin{eqnarray}}
\def\eea{\end{eqnarray}}

\begin{document}
\title{Probing the superconducting ground state of ZrIrSi: A $\mu$SR study}
\author{K. Panda}
\affiliation {Department of Physics, Ramakrishna Mission Vivekananda Educational and Research Institute, Howrah 711202, India}
\author{A. Bhattacharyya}
\email{amitava.bhattacharyya@rkmvu.ac.in}
\affiliation{Department of Physics, Ramakrishna Mission Vivekananda Educational and Research Institute, Howrah 711202, India} 
\author{D. T. Adroja} 
\affiliation{ISIS Facility, Rutherford Appleton Laboratory, Chilton, Didcot, Oxon, OX11 0QX, United Kingdom} 
\affiliation{Highly Correlated Matter Research Group, Physics Department, University of Johannesburg, Auckland Park 2006, South Africa}
\author{N. Kase}
\affiliation{Department of Applied Physics, Tokyo University of Science, Tokyo, 125-8585, Japan}
\author{P. K. Biswas}
\affiliation{ISIS Facility, Rutherford Appleton Laboratory, Chilton, Didcot, Oxon, OX11 0QX, United Kingdom}
\author{Surabhi Saha} 
\affiliation{Department of Physics, Indian Institute of Science, Bangalore, 560012, India}
\author{Tanmoy Das} 
\email{tnmydas@gmail.com}
\affiliation{Department of Physics, Indian Institute of Science, Bangalore, 560012, India}
\author{M. Lees}
\affiliation{Department of Physics, University of Warwick, Coventry CV4 7AL, United Kingdom}
\author{A. D. Hillier}
\affiliation{ISIS Facility, Rutherford Appleton Laboratory, Chilton, Didcot, Oxon, OX11 0QX, United Kingdom}

\begin{abstract}
The superconducting ground state of newly reported ZrIrSi is probed by means of $\mu$SR technique along with resistivity measurement. The occurrence of superconductivity at $T_\mathrm{C}$ = 1.7 K is confirmed by resistivity measurement. ZF-$\mu$SR study revealed that below $T_\mathrm{C}$, there is no spontaneous magnetic field in the superconducting state, indicates TRS is preserved in case of  ZrIrSi. From TF-$\mu$SR measurement, we have estimated the superfluid density as a function of temperature, which is described by an isotropic $s-$wave model with a superconducting gap $2\Delta(0)/k_\mathrm{B}T_\mathrm{C}$ = 5.1, indicates the presence of strong spin-orbit coupling. {\it Ab-initio} electronic structure calculation indicates that there are four bands passing through the Fermi level, forming four Fermi surface pockets. We find that the low-energy bands are dominated by the $4d$-orbitals of transition metal Zr, with substantially lesser weight from the $5d$-orbitals of the Ir-atoms.
\end{abstract}

\date{\today} 
\maketitle

\section{Introduction}
\noindent Recently, a number of studies have been carried out in ternary metal phosphide, silicide and arsenide with general formula  $TrT'X$ ($Tr$ and $T'$ are either 4$d$ or 3$d$ transition  elements and whereas $X$ is either a group IV or V member)~\cite{Barz1980,Muller1983, Seo1997, Ching2003, Keiber1984, Shirotani2001, Shirotani1998, Shirotani2000, Shirotani1999, Shirotani1995}. These systems have attracted considerable attention due to their relatively high superconducting transition temperature ($T_\mathrm{C}$), for example 15.5 K for hexagonal $h-$MoNiP~\cite{Shirotani2000}, 13 K for $h-$ZrRuP~\cite{Shirotani1993} and 12 K for $h-$ZrRuAs~\cite{Meisner1983}. These ternary equiatomic systems have provided a playground to investigate the role of spin orbit (SO) coupling in superconductivity, which were not so well studied in these systems. Compounds with Ir are often characterized with a strong SO coupling effect, due to presence of the Ir $5d$-orbitals. Superconductivity is observed in a number of Ir-based compounds such as Li$_\mathrm{2}$IrSi$_{3}$ ($T_\mathrm{C}$ = 4.2 K) ~\cite{Pyon2014, Lu2015}, IrGe ($T_\mathrm{C}$ = 4.7 K) ~\cite{Matthias1963}, RIr$_\mathrm{3}$ [$T_\mathrm{C}$ = 3.1 K (La), $T_\mathrm{C}$ = 3.3 K (Ce)] ~\cite{Hal, Sato2018,BhattacharyyaLaIr3},  CaIr$_\mathrm{2}$($T_\mathrm{C}$ = 5.8 K) ~\cite{Haldolaarachchige2015}, HfIrSi ($T_\mathrm{C}$ = 3.1 K) ~\cite{Kase2016} and ScIrP ($T_\mathrm{C}$ = 3.4 K) ~\cite{Pfannenschmidt}.  Cuamba {\it et.al.}~\cite{,Pyon2014, Lu2015, Matthias1963, Cuamba} suggests that the presence strong spin orbit (SO) coupling and a significant contribution to the total density of states (DOS) comes from the Ir-atom in most of the Ir-based compounds. Recently, we have reported time reversal symmetry (TRS) breaking superconductivity on the transition metal based caged type R$_\mathrm{5}$Rh$_\mathrm{6}$Sn$_\mathrm{18}$ (R = Lu, Sc, and Y) ~\cite{Bhattacharyya1,Bhattacharyya2,Bhattacharyya3} compounds due to  strong spin orbit coupling.

\noindent In ternary equiatomic compounds, superconductivity has only been found in two types of crystal structures: the first one is the hexagonal Fe$_\mathrm{2}$P-type (space group $P\bar{\mathrm{6}}m$2),~\cite{Shirotani2001,Shirotani2000, Ching2003} and the second one is the orthorhombic Co$_\mathrm{2}$Si-type (space group $Pnma$)~\cite{Shirotani1998,Shirotani2000, Ching2003, Muller1983}. It is interesting to note that in these systems $T_\mathrm{C}$ is strongly associated with crystal structure. Furthermore, $h-$Fe$_\mathrm{2}$P-type structure exhibit higher $T_\mathrm{C}$ than $o-$Co$_\mathrm{2}$Si-type structure, for example: $h-$ZrRuP  shows $T_\mathrm{C}$ at 13.0 K whereas $o-$ZrRuP exhibit $T_\mathrm{C}$ at 3.5 K. In case of $h-$Fe$_\mathrm{2}$P-type structure each layer is fiiled up with either $Tr$ and $T'$ or $X$ and $X$ elements. In case of $o-$ZrRuP,  Shirotani  {\it et al.}~\cite{Shirotani1999} reported the formation of two dimensional triangular  Ru$_\mathrm{3}$ clusters and in the basal plane they are connected through Ru-P ionic bonds. It also connected through Zr-Ru bonds where Zr atoms occupy the $ z=1/2$ plane. If phosphorus is replaced by the more electronegative silicon then nearest neighbour Ru-Ru bond length enlarged to 2.87 \AA. Surprisingly $o-$MoRuP shows superconductivity at 15.5 K which is as high as isoelectronic $h-$ZrRuP ($T_\mathrm{C}$ = 13 K)  and $h-$MoNiP ($T_\mathrm{C}$ = 13 K).  Ching {\it  et. al.}~\cite{Ching2003} shown that in $o-$MoRuP and $o-$ZrRuP higher value of DOS at the FL is directly related to higher $T_\mathrm{C}$, as suggested for BCS superconductor. In these systems, the density of states are governed by Mo-$4d$ orbitals. Ching {\it  et. al.}~\cite{Ching2003} have calculated the values of density of states which are 0.46 states per eV atom and 0.33 states per eV atom for $o-$MoRuP and $o-$ZrRuP, respectively.

\begin{figure*}[t]
\centering
   \includegraphics[height=0.3\linewidth,width=\linewidth]{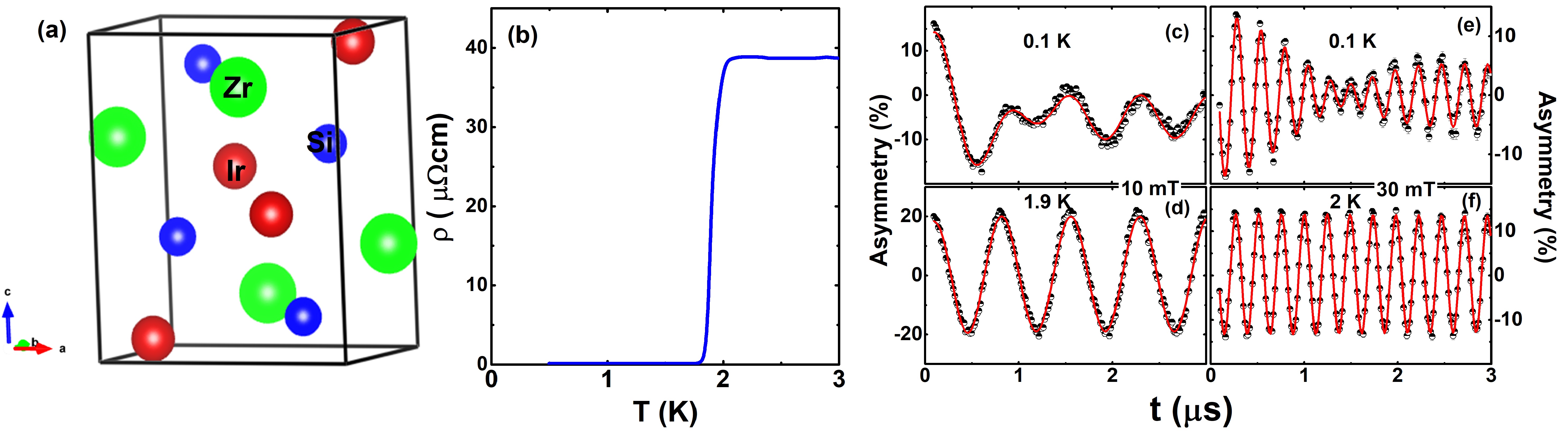}\hfil
\caption{(a) The crystal structure of ZrIrSi which crystallize in the orthorhombic structure with space group $Pnma$, where green, red and blue dark yellow symbol represents Zr (big in size), Ir (medium), and Si atoms (small) respectively. (b) Temperature variation of resistivity in zero field. Time dependence TF-$\mu$SR asymmetry spectra for ZrIrSi recorded at (c) $T$ = 0.1 K (d) $T$ = 1.9 K in presence of an applied field $H$ = 10 mT and  at (e) $T$ = 0.1 K at (f) $T$ = 2.0 K in an applied field $H$ = 30 mT. The red solid line shows the fittings to the data using Eqn. 1 described in the text.}
\label{physical properties}
\end{figure*}

\noindent Therefore, to investigate the superconducting pairing mechanism in ZrIrSi, we have reported a systematic $\mu$SR studies. Zero field (ZF)-$\mu$SR is a powerful technique to know the time reversal symmetry (TRS) breaking in the superconducting state~\cite{Sonier}. ZF-$\mu$SR data indicates the absence of any spontaneous magnetic fields below $T_\mathrm{C}$, thus implying that TRS is not broken in the superconducting state. The superfluid density as a function of temperature is determined from the depolarization rate of the transverse field (TF)-$\mu$SR which is well described by an isotropic $s-$wave model. These results are further supported by {\it ab-initio} electronic structure calculation.

\section{Experimental Details}

\noindent For this study, a polycrystalline sample of ZrIrSi was synthesized using typical arc melting process on a water-cooled copper hearth using Zr (99.99\%), Ir (99.99\%), and Si (99.999\%) in a stoichiometric ratio. The arc melted ingot was remelted several times to confirm the phase homogeneity. After that the sample was annealed at 1000$^{0}$C for a week in a sealed vacuum quartz tube. X-ray diffraction were carried out using Cu-$K_{\mathrm{\alpha}}$ radiation. Electrical resistivity measurement was done in a standard dc-four probe technique down to 0.5 K.

\noindent $\mu$SR experiments were performed at the ISIS pulsed neutron and muon source of Rutherford Appleton Laboratory, UK using MuSR spectrometer with 64 detectors at transverse and longitudinal direction~\cite{Lee1999}. 100\% spin-polarized muons were probed into the sample as a result positive muons decomposed into a positron, preferably in the direction of muon spin vector and two neutrinos with an average lifetime of 2.2 $\mu$s. These positrons are detected by the detectors, placed either forward ($F$), or backward ($B$) direction. The time dependence of $\mu$SR asymmetry spectra $A$ is calculated as $A(t)=\frac{N_{\mathrm{F}}(t)-\alpha N_{\mathrm{B}}(t)}{N_{\mathrm{F}}(t)+\alpha N_{\mathrm{B}}(t)}$, where $N_{\mathrm{F}}(t)$ and $N_{\mathrm{B}}(t)$ are the number of positrons counted in the forward and  backward detectors respectively, and $\alpha$ is an instrumental calibration factor. ZF-$\mu$SR is carried out with detectors in the longitudinal configuration, where a correction coil is applied to neutralize any stray magnetic fields up to 10$^{-4}$~mT. ZF-$\mu$SR measurement is crucial to understand the type of pairing symmetry in superconductors \cite{Sonier}. TF-$\mu$SR measurements were carried out in the vortex state in presence of 10, 20, 30 and 40 mT applied field, which is above the lower critical field $H_\mathrm{c1}$ (= 0.7 mT),  and below the upper critical field $H_\mathrm{c2}$(= 0.6 T).  The sample was mounted onto a high purity (99.995\%) silver sample holder using diluted GE-varnish and then wrapped with thin silver foil. This was inserted in the sample chamber using a dilution refrigerator that can go down to 50 mK. We have analyzed the $\mu$SR data using WiMDA~\cite{Pratt2000} software.
 
\begin{figure*}[t]
\centering
    \includegraphics[height=0.3\linewidth,width=\linewidth]{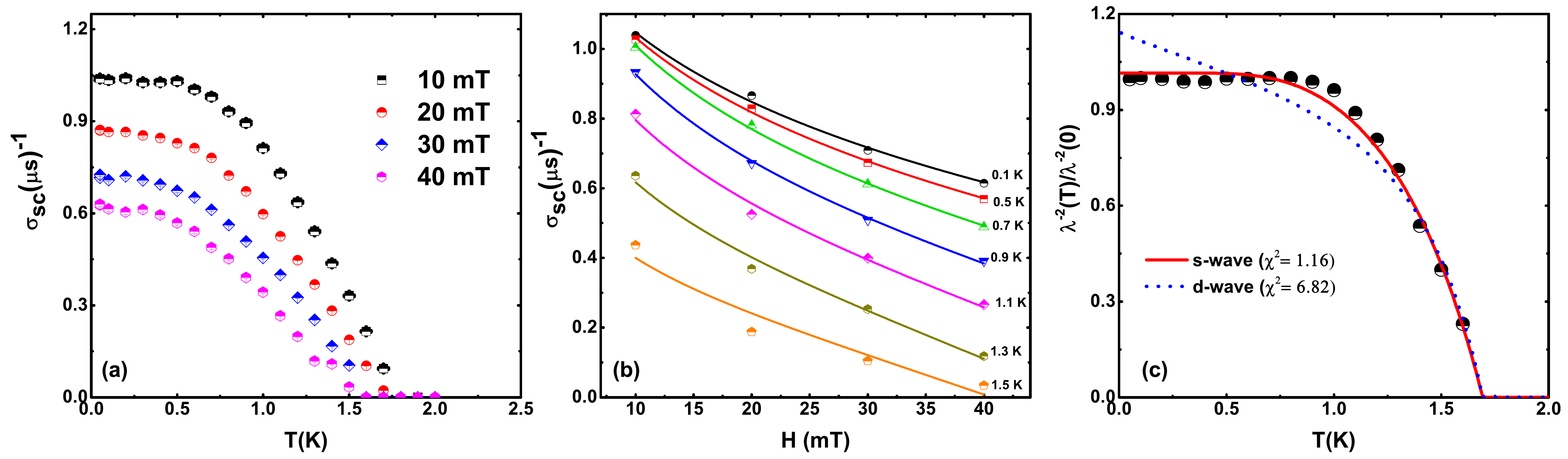}
    \caption{(a) The superconducting depolarization rate $\sigma_\mathrm{sc}(T)$ as a function of temperature in presence of  applied field of 10 mT$\leq$ H$\leq$ 40 mT. (b) Magnetic field dependence of the muon spin-depolarization rate is shown for a range of different temperatures. The solid lines are the results of fitting the data using Brandt equation as discussed in the text Eq. \ref{eqn2}. (c) The inverse magnetic penetration depth squared as a function of temperature is shown here. The lines show the fits using $s-$wave (solid), and $d-$wave (dashed) gap functions.}
\label{musrdata}
\end{figure*}

\section{Results and discussion}

\subsection{Crystal structure and resistivity}

\noindent X-ray powder diffraction data revealed that ZrIrSi crystallize in the orthorhombic structure (space group $Pnma$) as displayed in Fig.~\ref{physical properties}(a). The calculated lattice parameters are $a$ = 6.557~\AA, ~ $b$ = 3.942~\AA ~ and $c$ = 7.413~\AA, which are in agreement with previous report~\cite{Kase2016}. The temperature ($T$) variation of the electrical resistivity $\rho (T)$ in zero applied magnetic field is presented in Fig. \ref{physical properties}(b). The electrical resistivity data reveals superconductivity at $T_{\mathrm{C}}$ = 1.7 K. Kase {\it et. al.}~\cite{Kase2016} have estimated the  Ginzburg Landau coherence length ($\zeta$) = 23.1 nm. It is interesting to note that $T-$ dependence of upper critical field shows a convex curvature~\cite{Kase2016}, which might suggest the presence of SO coupling. Similar curvature is also found in R$_\mathrm{3}$T$_\mathrm{4}$Sn$_\mathrm{13}$ (R = La, Sr; T = Rh, Ir) which is a SO coupled superconductor~\cite{Kase2011}.

\subsection{TF-$\mu$SR analysis}

\noindent To explore the pairing mechanism and gap structure of superconducting state of ZrIrSi, TF-$\mu$SR measurements were performed down to 0.05 K. Fig.~1(c)-(f) represents the TF-$\mu$SR asymmetry time spectra in presence of 10 mT and 30 mT applied magnetic field at above and below $T_\mathrm{C}$.  Below $T_\mathrm{C}$ the spectra depolarizes strongly because of the inhomogeneous field distribution in the vortex state. TF-$\mu$SR data fitted using two Gaussian oscillatory, functions~\cite{Bhattacharyyarev, AdrojaThFeAsN, BhattacharyyaThCoC2}:

\begin{equation}
G_\mathrm{TF}(t) = \sum_{i=1}^{2} A_\mathrm{i}\cos(\omega_\mathrm{i}t+\phi)\exp(-\frac{\sigma_\mathrm{i}^{2}t^{2}}{2})
\end{equation}

where $A_\mathrm{i}$, $\sigma_\mathrm{i}$, $\omega_\mathrm{i}$, $\phi$ is the initial asymmetry, Gaussian relaxation rate, muon spin precession frequency and the initial phase of the offset, respectively. In this fit $\sigma_\mathrm{i}$  for the $2^{nd}$ part is equal to zero, which corresponds to the background term. This term comes from those muons which missed the sample and directly hit the silver sample holder and therefore the depolarization of this oscillating term is zero, i.e. $\sigma_\mathrm{2}$ = 0 as silver has a minimal nuclear moment. $\sigma_\mathrm{1}$ can be expressed as $\sigma_\mathrm{1} = \sqrt{\sigma_\mathrm{sc}^{2}+\sigma_\mathrm{n}^2}$, where $\sigma_\mathrm{sc}$ comes from superconducting part and $\sigma_\mathrm{n}$ comes from nuclear magnetic dipolar moment which is fixed in the entire temperature range, supported by the ZF-$\mu$SR later. \\

\noindent The temperature variation of $\sigma_\mathrm{sc}$ is depicted in Fig.~\ref{musrdata}(a). As $H_\mathrm{c2}$ value is low in this sample, $\sigma_{\mathrm{sc}}$ depends on the applied field as displayed in Fig.~\ref{musrdata}(b). Brandt~\cite{Brandt2003, Brandt1988} has reported that for a superconductor with $H_\mathrm{ext}/H_\mathrm{c2}$ $\leq$0.25, $\sigma_\mathrm{sc}$ is associated with the London penetration depth [$\lambda(T)$] by the following equation:

\begin{equation}
\begin{split}
\sigma_\mathrm{sc}[\mu s^{-1}] = 4.83 \times 10^{4}(1-H_\mathrm{ext}/H_\mathrm{c2}) \\ 
\times [1+1.21(1-\sqrt{{(H_\mathrm{ext}/H_\mathrm{c2}})} )^{3}]\lambda^{-2}[nm]
\end{split}
\label{eqn2}
\end{equation}

This equation is a good approximation for $\kappa \geq 5$, which is valid for our case as $\kappa = 40.5$ for ZrIrSi~\cite{Kase2016}. From this relation we have determined the temperature dependence of $\lambda(T)$ and $\mu_{0}H_\mathrm{c2}(T)$. Isothermal cuts perpendicular to the temperature axis of $\sigma_{\mathrm{sc}}$ data sets were used to determine the $H$-dependence of the depolarization rate $\sigma_{\mathrm{sc}}(H)$ as displayed in Fig.\ref{musrdata}(b). We have estimated the London penetration depths $\lambda$ = 254.4(3) nm, using $s$-wave model.

\noindent We have plotted the temperature variation of normalized $\lambda^{-2}(T)/\lambda^{-2}(0)$, which is directly proportional to the superfluid density. $\lambda^{-2}(T)/\lambda^{-2}(0)$ data were fitted using the following equation~\cite{Prozorov, BhattacharyyaBiS2, AdrojaK2Cr3As3, AdrojaCs2Cr3As3, DasLaPt2Si2}:

\begin{eqnarray}
\frac{\sigma_{sc}(T)}{\sigma_{sc}(0)} &=& \frac{\lambda^{-2}(T)}{\lambda^{-2}(0)}\\
 &=& 1 + \frac{1}{\pi}\int_{0}^{2\pi}\int_{\Delta(T)}^{\infty}(\frac{\delta f}{\delta E}) \times \frac{EdEd\phi}{\sqrt{E^{2}-\Delta(T})^2} \nonumber
\end{eqnarray}

here $f$ is the Fermi function which can be expressed as $f= [1+\exp(-E/k_\mathrm{B}T)]^{-1}$. $\Delta(T,0) = \Delta_{0}\delta(T/T_\mathrm{C})g(\phi)$ whereas $g(\phi)$ is the angular dependence of the gap function, $\phi$ is the azimuthal angel in the direction of FS. The temperature variation of the superconducting gap is approximated by the relation $\delta(T/T_\mathrm{C}) = \tanh \{{1.82[1.018 (T_\mathrm{C}/T -1)]^{0.51}} \}$. The spatial dependence $g(\phi$) is substituted by (a) 1 for $s-$wave gap, (b) $\vert\cos(2\phi)\vert$ for $d-$wave gap with line nodes.

\noindent Fig.~\ref{musrdata}(c) represents the fits to the $\lambda^{-2}(T)/\lambda^{-2}(0)$ data of ZrIrSi  using a single gap $s-$wave and nodal $d-$wave models. It is clear that the data can be well described by the isotropic $s-$wave model with a gap value 0.374 meV. This model gives a gap to $T_\mathrm{C}$ ratio, $2\Delta(0)/k_\mathrm{B}T_\mathrm{C}$ = 5.10. The higher value of gap compare to BCS gap (3.53) suggest the presence of strong SO coupling. Similar high gap value was obtained for Ir-based superconductors, for example: IrGe [$2\Delta(0)/k_\mathrm{B}T_\mathrm{C}$ = 5.14]~\cite{Cuamba, Matthias1963}, CaIrSi$_\mathrm{3}$  [$2\Delta(0)/k_\mathrm{B}T_\mathrm{C}$ = 5.4]~\cite{SinghCaTSi3}. On the other hand $d-$wave model is clearly not suitable for this system as the $\chi^{2}$ value increased significantly for this fit ($\chi^{2}$ = 6.82). As ZrIrSi is a type II superconductor, supposing that approximately all the normal states carriers ($n_\mathrm{e}$) contribute to the superconductivity ($n_\mathrm{s} \approx$ n$_{e}$), superconducting carrier density $n_\mathrm{s}$, the effective-mass enhancement $m^{*}$ have been estimated to be $n_\mathrm{s}$ = 6.9(1) $\times$ 10$^{26}$ carriers $m^{-3}$, and $m^{*}$ = 1.474(3) $m_\mathrm{e}$ respectively for ZrIrSi. Detail calculations can be found in Ref.[42-44] ~\cite{Hillier1997,Adroja2005, AnandLaIrSi3}.
\begin{figure}[t]
\centering
    \includegraphics[width=0.9\linewidth]{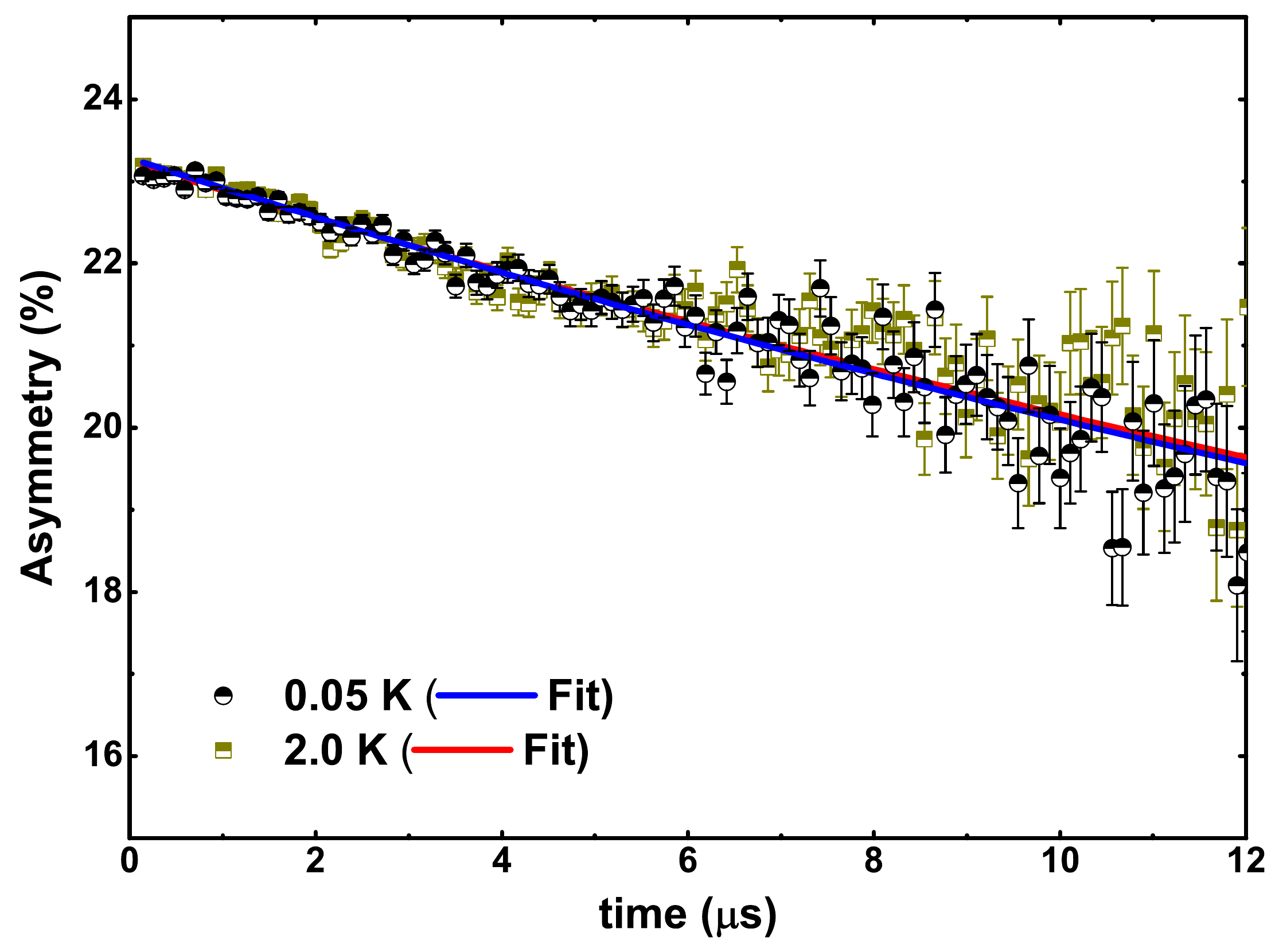}
\caption{ZF-$\mu$SR asymmetry time spectra for ZrIrSi at  0.05 K (black circle) and 2 K (dark yellow square) are shown together.  The lines are the least squares fit to the data using Eq. 4.}
\label{ZF}
\end{figure}

\subsection{ZF-$\mu$SR analysis}

\noindent In order to investigate the pairing mechanism in the superconducting ground state, we used ZF-$\mu$SR study. The time evolution of asymmetry spectra is shown in Fig.~\ref{ZF} for $T$ = 0.05 K $< T_\mathrm{C}$ and $T$ = 2 K $> T_\mathrm{C}$. The spectra below and above $T_\mathrm{C}$ are found identical, ruling out the presence of any magnetic field. This reveals that the TRS is preserved in the superconducting state of ZrIrSi. This ZF data were fitted by a Lorentzian function: 

\begin{equation}
G_\mathrm{ZF}(t) = A_\mathrm{0}(t)\exp{(-\lambda t)}+A_\mathrm{bg}
\end{equation} 

\noindent $A_\mathrm{0}$ and $A_\mathrm{bg}$ are the sample and background asymmetry respectively, which are nearly temperature independent. $\lambda$ is the relaxation rate which comes from nuclear moments. The red and blue lines in Fig.~\ref{ZF} indicate the fits to the ZF-$\mu$SR data. The fitting parameters of the ZF-$\mu$SR asymmetry data are as follows: $\lambda$ = 0.030(9) $\mu \mathrm{s}^{-1}$ at  0.05 K  and $\lambda$ = 0.026(3) $\mu \mathrm{s}^{-1}$ at 2 K.  The change is the relaxation rate is within the error bar, indicates no clear evidence of TRS breaking in ZrIrSi.

\subsection{Theoretical Calculations}
ZrIrSi unit-cell has a $\it{mmm}$ point group symmetry and it belongs to the $\it{Pnma(62)}$ space group (orthorhombic crystal structure). We have used the Vienna Ab-initio Simulation Package(VASP) for ab-initio electronic structure calculation. The projector augmented wave (PAW) pseudo-potentials are used to describe the core electrons and for the exchange-correlation functional Perdew-Burke-Ernzerhof (PBE) form is used. We have used a local density application (LDA) functional with a cut-off energy for the plane wave basis set of 500 eV.  The Monkhorst-Pack $k$-mesh is set to $14\times14\times14$ in the Brillouin zone for the self-consistent calculation. The relaxed lattice parameters were found to be as follows $a$ = 3.9643~\AA, $b$ = 6.5893~\AA, and $c$ = 7.4070~\AA~and $\alpha=\beta=\gamma=90\degree$. To deal with the strong correlation effect of the $d$-electrons of the Ir atoms, we employed the LDA+U method with $U$ = 2.8 eV. For the Fermi surface calculation, we have used a larger $k$-mesh of $31\times31\times31$.

\begin{figure}[t]
\centering
   \includegraphics[width=0.99\linewidth]{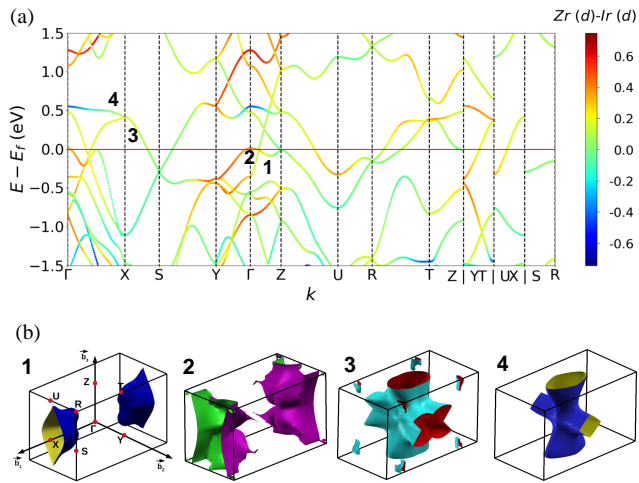}
    \caption{(a) Computed DFT band structure is plotted along the standard high-symmetric directions for the orthorhombic crystal structure of ZrIrSi. The bands are colored with the difference in the $d$-orbital weight of the Zr and Ir-atoms, where red color gives stronger Zr-$d$ orbital weight while blue color dictates Ir-$d$ orbital weight. Four bands which pass through the Fermi level are indicated by 1,2,3,4 numbers. (b) Corresponding Fermi surfaces are plotted in the full three-dimensional Brillouin zone. We find that Fermi surface 1 and 2 form pockets around the X-point, while Fermi surface 3 and 4 constitute pockets centering the $\Gamma$-point. The colors on the Fermi surface only aid visualization.}
\label{DFT}
\end{figure}

In Fig. \ref{DFT}, we show the band structure and Fermi surface plots.  We find that there are four bands passing through the Fermi level, forming four Fermi surface pockets. Two Fermi pockets are centered around the $\Gamma$-point, while the other two pockets are centered around the X-point. Unlike the multi-gap superconductivity in MgB$_2$ \cite{MgB2} and Mo$_8$Ga$_{41}$\cite{Mo8Ga41} which are driven by the presence of multiple Fermi surfaces, we do not find any evidence of multigap superconductivity in ZrIrSi, which is in agreement with the TF-$\mu$SR results. This is presumably because of the absence of $E_{2g}$-phonon mode which could enable inter-band scattering, while in the present case phonon modes cause intra-band electron-phonon coupling. Moreover, we observe substantial three-dimensionality in all four Fermi surfaces. This substantially weakens Fermi surface nesting strength. Thus the possibility of an inter-band nesting driven unconventional $s^{\pm}$-pairing symmetry is suppressed as compared to a two-dimensional iron-pnictide family with similar Fermi surface topology (see, e.g., Ref.~\cite{FeAs}).

Finally, we study the orbitals' contributions to the low-energy electronic states. We find that the low-energy bands are dominated by the $4d$-orbitals of transition metal Zr, with substantially lesser weight from the $5d$-orbitals of the Ir-atoms. Due to the presence of $d$-orbitals in the low-energy states, it is natural to anticipate the involvement of strong correlation in these systems and hence strong coupling superconductivity which changes from typical electron-phonon to quasiparticle-phonon mechanism within the Eliashberg theory\cite{QPCoupling}. However, to our surprise, we find a substantially lower effective mass $1.5 m_e$ (where $m_e$ is the bare electron's mass) which is in remarkable agreement with the results from the TF-$\mu$SR measurements. This is well captured within our LDA+U calculation without essentially including dynamical correlations. We repeated the calculations for isostructural compounds TiIrSi and HfIrSi and find that the essential Fermi surface topology and three-dimensionality remain the same in all three materials (not shown). Therefore, we conclude that the superconductivity in ZrIrSi and its isostructural materials (such as TiIrSi and HfIrSi) can be well understood within the conventional BCS theory. Although our estimates of the BCS ratio of 5.1 is slightly higher than the BCS estimate of 3.5, however, we believe this slight increment is caused by the spin-orbit coupling of the Ir-atoms, and the Fermi surface anisotropy.        

\section*{}

\section{Conclusions}

\noindent In conclusion, we have performed ZF and TF-$\mu$SR measuremnts in the mixed state of ZrIrSi. Using Brandt's equation we have determine the temperature dependence of the magnetic penetration depth. The superfluid density $n_\mathrm{s} \propto 1/\lambda^{2}$ well described by an isotropic $s-$wave model. The obtained gap value is $2\Delta(0)/k_\mathrm{B}T_\mathrm{C}$ = 5.1, which suggest ZrIrSi to be a strongly coupled BCS superconductor. {\it Ab-initio} electronic structure calculation indicates BCS superconductivity, which supports our experimental results. The low-energy bands are dominated by the $4d-$orbitals of the transition metal Zr, with a substantially lesser weight from the $5d-$orbitals of the Ir-atoms. ZF-$\mu$SR reveals, there is no spontaneous magnetic field below $T_\mathrm{C}$, which suggest the absence of TRS breaking. The present results pave the way to develop a realistic theoretical model to interpret the origin of superconductivity in ternary systems.

\subsection*{Acknowledgements}{ KP acknowledge the financial support from DST India, for Inspire Fellowship (IF170620). AB would like to acknowledge DST India, for Inspire Faculty Research Grant (DST/INSPIRE/04/2015/000169), and UK-India Newton funding for funding support. DTA and ADH would like to thank the Royal Society of London for the UK-China Newton funding and CMPC-STFC, grant number CMPC-09108, for financial support. TD acknowledges the financial support from the Science and Engineering Research Board (SERB), Department of Science \& Technology (DST), Govt. of India for the Start-Up Research Grant (Young Scientist).}

\end{document}